\documentclass[usegraphicx,useAMS,usenatbib]{mn2e}

\def\iron{\hbox{[Fe\,{\sc ii}]}}
\def\nickel{\hbox{[Ni\,{\sc ii}]}}

\def\hi{\hbox{H\,{\sc i}}}
\def\hei{\hbox{He\,{\sc i}}}
\def\sivi{\hbox{[Si\,{\sc vi}]}}

\def\alix{\hbox{[Al\,{\sc ix}]}}

\title[SINFONI spectroscopy of Super-antennae]{VLT/SINFONI Integral Field Spectroscopy of The Super-antennae\thanks{Based on observations collected at the European Southern Observatory, Paranal, Chile (60.A-9041(A)).}}
\author[Reunanen et al.]{J. Reunanen$^{1}$, 
L. E. Tacconi-Garman$^{2}$, 
V. D. Ivanov$^{3}$\\
$^{1}$Leiden Observatory, PO Box 9513, 2300 RA Leiden, The Netherlands; reunanen@strw.leidenuniv.nl\\
$^{2}$European Southern Observatory, Karl Schwarzschildstrasse 2, 85748 Garching, Germany; ltacconi@eso.org\\
$^{3}$European Southern Observatory, Alonso de Cordova 3107, Santiago 19, Chile; vivanov@eso.org\\}
\begin{document}

\date{}

\pagerange{\pageref{firstpage}--\pageref{lastpage}} \pubyear{2007}

\maketitle

\label{firstpage}

\begin{abstract}
We present the results of $H$- and $K$-band VLT/SINFONI integral field
spectroscopy of the ULIRG IRAS 19254$-$7245 (The Super-antennae), an
interacting double galaxy system containing an embedded AGN\@. Deep
$K$-band spectroscopy reveals Pa$\alpha$ arising in a warped disc with
position angle of 330{\degr} and an inclination $i=40-55${\degr}. The kinemetric
parameters derived for H$_2$ are similar to Pa$\alpha$. Two
high-ionization emission lines, {\sivi} and {\alix}, are detected and we
identify as {\nickel} the line observed at 1.94$\,\mu$m.  Diluting
non-stellar continuum, which was previously detected, has decayed, and
the $H$-band continuum emission is consistent with pure stellar
emission. 
Based on H$_2$ emission line ratios it is likely that at the central
1-kpc region H$_2$ is excited by UV fluorescence in dense clouds while shock
excitation is dominant further out. This scenario is supported by very
low Pa$\alpha$ to H$_2$ line ratio detected outside the nuclear
region and non-thermal ortho/para ratios ($\sim$2.0 -- 2.5) close to the nucleus.

\end{abstract}


\section{Introduction}

The Super-antennae (IRAS 19254$-$7245) at a redshift of $z\simeq0.062$ (distance of
$\sim260$ Mpc assuming $H_0$ = 71 km s$^{-1}$ Mpc$^{-1}$; Spergel et al.~2003) is
an interacting double galaxy system where the two components are separated by
$\sim$8\arcsec ($\sim$2 kpc).  Visually the most notable feature of the system are
the thin antennae extending up to 5\arcmin ($\sim$350 kpc), far longer
than similar feature of the Antennae galaxy ($\sim$100 kpc).

The starburst activity powering the total infrared luminosity of
$10^{11.91}$ L$_{IR}$/L$_\odot$ (Duc, Mirabel \& Maza 1997) in the
system has been triggered by a 3 : 1 mass ratio encounter (Dasyra et
al.~2006). The northern nucleus is less luminous than the southern
nucleus in the infrared, and is believed to be in a post-starburst
stage (e.g.~Berta et al.~2003), while the southern galaxy is a
starburst with an embedded AGN\@. Berta et al.~(2003) found the
spectrophotometric characteristics of the southern nuclei to be mostly
consistent with two main stellar populations: an intermediate aged (1\,Gy)
starburst, which represents $\sim$35 percent of the luminous mass of the galaxy
and an old 12\,Gy population representing the remaining $\sim$65 percent
of the total
luminous mass. In addition, the southern nucleus has an intense ongoing
starburst, which contributes little to the total luminous mass.

The evidence for the presence of an AGN comes from the optical emission line
ratios (Mirabel, Lutz \& Maza 1991), IR coronal-emission lines (Vanzi
et al.~2002),  observations of a hard, flat X-ray component above 2
keV (Pappa, Georgantopoulos \& Stewart 2000), and strong Fe K$\alpha$
emission lines (Braito et al.~2003) interpreted as a Compton-thick
X-ray source. Like in other LIRGs/ULIRGs, a long-standing question has
been whether the spectral energy distribution of the southern galaxy
is dominated by the starburst or the AGN\@.  The mid-IR spectroscopic
diagnostics are intermediate between those for starbursts and AGN (Genzel et
al.~1998). Recently, Charmandaris et al.~(2002) and Risaliti et
al.~(2003; 2006) claimed the AGN is the main energy source in the southern
galaxy based on mid-IR ISO and deep $L$-band ground-based
spectroscopy, respectively.

Based on the integral field spectroscopic data obtained with
VLT/SINFONI on the southern galaxy, we discuss the nuclear spectrum,
excitation of molecular H$_2$ emission lines and the morphology and
kinemetric properties of ionized and molecular gas.

In addition to the high-excitation emission lines requiring the
presence of either strong ionising continuum or fast shocks with $v_s
\simeq 200$ km s$^{-1}$ or greater (Contini \& Viegas 2001), other AGN
indicators in near-IR wavelength range covered in this work are the variability
of the emission lines and non-stellar continuum. Considering the
distance and brightness of the Super-antennae, supernovae and their
subsequent evolution are unlikely to produce detectable spectroscopic
variability at the nucleus.

The dominant excitation mechanism for H$_2$ has been under debate for
years, the two main candidates being UV fluorescence (e.g.~Black \&
van~Dishoeck 1987; Sternberg \& Dalgarno 1989) and shocks
(e.g.~Hollenbach \& McKee 1989). The interpretation of H$_2$ line
ratios is complicated due to the strong dependence of line ratios on
density; the energy levels are thermalised through inter-molecule
collisions and the line ratios in dense clouds illuminated by intense
UV radiation are very similar to those from purely thermal excitation,
either through shocks or X-rays.  Furthermore the H$_2$ line ratios
are affected by the spin degeneracy of the levels. Radiative decay
between ortho (odd $J$) and para (even $J$) is not possible due to
different spin and in the ground electric state only $\Delta J$ = 0,
$\pm$2 transitions are possible. For purely thermal excitation the
ratio between the ortho and para levels is 3.0, but is lower for
fluorescence in tenuous gas ($\sim$1.9 using the 1--0 S(1), S(2) and
S(3) transitions; Black \& van~Dishoeck 1987).

As the Super-antennae is an interacting system, the kinematics is
expected to be more complex than in the case of non-interacting
galaxies. In the literature multiple gaussian have been used to
approximate optical emission lines (Vanzi et al.~2002; Colina, Lipari
\& Macchetto 1991). Interpretating the results of multi-gaussian fits
in a physically meaningful manner is however non-trivial, unless the
components are clearly separated from each other. A more qualitative
approach for tracing multiple velocity components is using the
Gauss-Hermite series to describe the line profile (van der Marel \&
Franx 1993; Gerhard 1993). Strong kinematical components can also be
revealed by fitting a warped disc model to the observed velocities.

The paper is organised as follows. In Section 2, the
observations and data reduction are described. In Section 3 the
results are presented and in Section 4 we summarise the conclusions of
the work.

\section{Observations and data reduction}

The southern galaxy of the Super-antennae was observed during SINFONI
(Spectrograph for INtegral Field Observations in the Near Infrared;
Eisenhauer et al.~2003; Bonnet et al.~2004) Science Verification in
Paranal, Chile, in August 2004 both in $H$ ($R \simeq 2900$) and $K$ bands
($R\simeq 4500$) with a 2048$^2$ pixel engineering grade detector. This
engineering grade detector had two significant defects: a
``glowcenter'' at 2.38$\,\mu$m and a large dead region at $>$2.4$\,\mu$m.
Unfortunately these artifacts compromised the data quality at the
nucleus. The 250 mas pixel scale was used during the observations without
Adaptive Optics providing a field of view (f.o.v) of
8\arcsec$\times$8\arcsec. The total integration time was 14000~s in $K$ and
1800~s in $H$-band. The average $V$-band seeing measured by the
Differential Image Motion Monitor (DIMM) during the observations was
0\farcs94 and 0\farcs77 during the $K$-band and $H$-band observations, respectively, and the conditions
were non-photometric. The observations were done following the
classical OSSO sequence of alternating object and sky exposures.

The data reduction was done with the \textit{spred} software package
(Schreiber at al.~2004) developed at the  Max Planck Institut f\"ur 
Extraterrestrische
Physik. 
The data were bad pixel and distortion corrected and sky
subtracted using the nearest sky exposure.  Instrumental flexure was
tracked by cross-correlating the night sky OH emission lines before sky
subtraction, and wavelength calibration was done by fitting a
quadratic polynomial to Ne arc lamp frames for the $K$-band and Xe+Ar
frames the  $H$-band. Due to unstable conditions during the
observations, the night sky emission lines and thermal background at
the red end of the $K$ band were not perfectly removed. This residual
background was removed by calculating a median value along the edges
of the cubes at each spectral plane and subtracting it. While this
subtraction unavoidably removes faint continuum emission from the
galaxy, the emission lines are unaffected as not even Pa$\alpha$ can
be detected this far out and is not present in the subtracted spectra.

The relative wavelength shifts caused by the instrumental flexure
between different exposures were determined by cross-correlation after
collapsing the reconstructed data cubes. The cubes were combined and
divided by the telluric standard spectra (Hip099481; B9.5V class star)
in order to remove atmospheric signatures. Finally, the flux
calibration was done by comparison with broad-band images (Vanzi et
al.~2002). We estimate the absolute flux calibration is accurate to
within 5--10 percent.

In order to enhance the signal-to-noise ratio of the images the data
were smoothed with a 3 pixel $\times$ 3 pixel boxcar (0\farcs38 $\times$
0\farcs38), which is much smaller than the resolution of the data.
Emission line measurements were done automatically with scripts
written by us for IRAF\footnote{IRAF is distributed by the National
Optical Astronomy Observatories, which are operated by the Association
of Universities for Research in Astronomy, Inc., under cooperative
agreement with the National Science Foundation.}. If a detected line 
was significantly narrower than the instrumental profile,
it was subtracted from the spectra and the fit was repeated.
Finally, the detections were inspected and all obviously
wrong fits due to cosmic ray residuals or noise were rejected. In the case
of Pa$\alpha$, which has multiple components that are difficult to fit
automatically, the fit was carried over interactively within
the central region.  In all the images shown in later sections the
lowest level displayed is at 2.5$\sigma$ unless otherwise stated.

\section{Results}

\begin{figure}
\includegraphics[width=8cm]{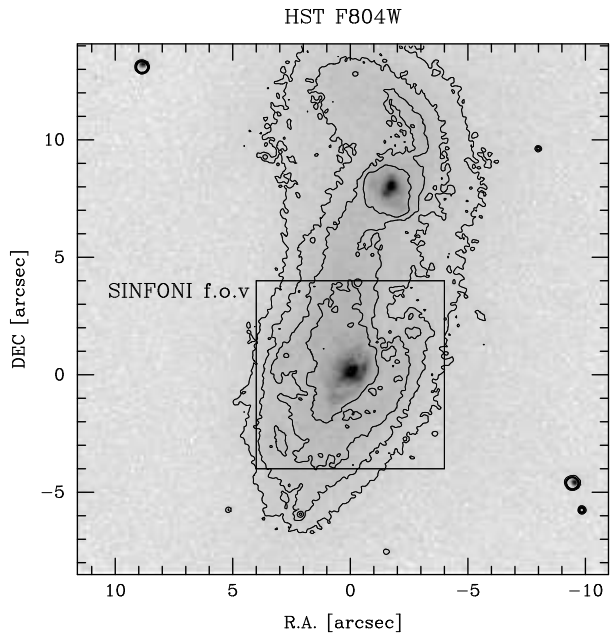}
\caption{HST F804W filter image of the Super-antennae.  The SINFONI
field of view is indicated with a box. The antennae of the galaxy do
not fit within the area depicted in this figure. East is left and
north is up in this and any subsequent images.}
\label{fig_f804w}
\end{figure}

A broadband F804W image of the Super-antennae obtained from the HST
data archive is presented in Fig. \ref{fig_f804w}. The SINFONI field was
centred on the southern nucleus and the field of view contains part
of the bridge detected in broadband imaging connecting the two
nuclei. 

\subsection{Nuclear spectrum}

\begin{figure*}
\includegraphics[width=16cm]{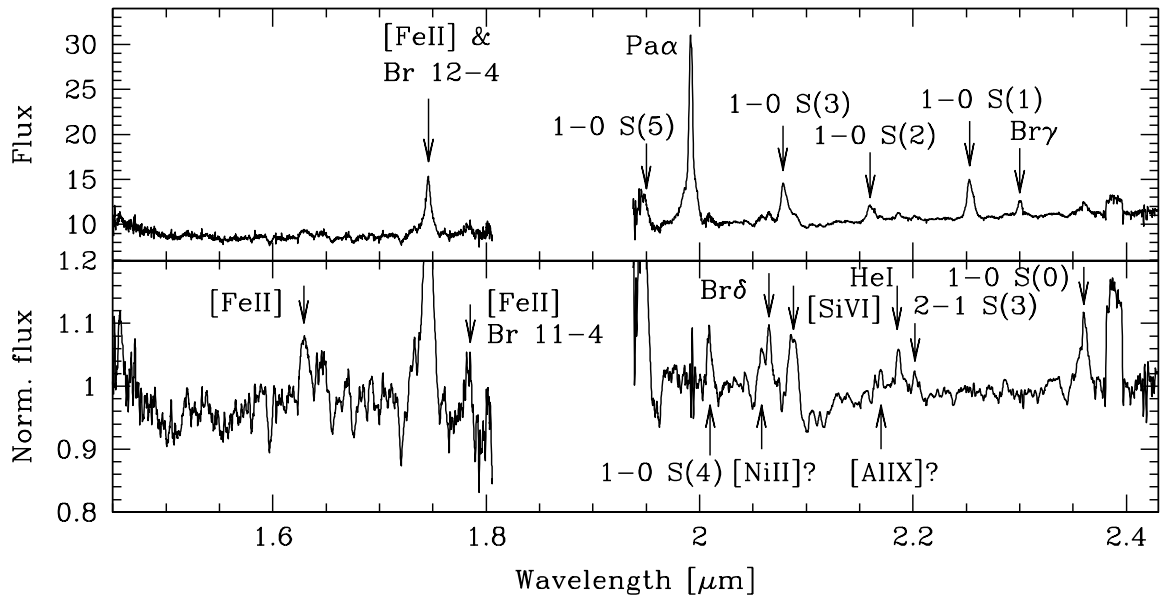}
\caption{The summed spectrum of the central 5$\times$5 spectra
(0\farcs6$\times$0\farcs6) in units
of 10$^{-13}$ ergs cm$^{-1}$ s$^{-1}$ $\mu$m$^{-1}$ with line identifications
as indicated ({\em upper
panel}). In the {\em lower panel} the normalised residuals after
subtracting the strongest {\hi} (Pa$\alpha$, Br$\gamma$) and H$_2$
lines (1--0 S(1), S(2), S(3)) and smoothing with residual with five
pixel boxcar are shown. The x-axis represents the observed wavelength.
\label{fig_nuc}}
\end{figure*}

The nuclear spectrum of the Super-antennae is displayed in
Fig. \ref{fig_nuc}. In addition to Pa$\alpha$ fortuitously located at an
atmospherically relatively clean region at $\sim$2$\,\mu$m, the $K$-band
spectrum shows many prominent molecular H$_2$ lines, hydrogen
recombination lines Br$\gamma$ and Br$\delta$ and {\hei} 2.058$\,\mu$m. 
A careful study reveals also two coronal lines - {\sivi} 1.962$\,\mu$m and 
{\alix} 2.04$\,\mu$m 
(redshifted to 2.0833$\,\mu$m and 2.166$\,\mu$m, respectively). 
We have compared the observed fluxes
to the ones reported by Vanzi et al.~(2002) in identical
apertures. While the H$_2$ line ratios are in good agreement with
Vanzi et al.~(2002), the hydrogen recombination lines Br$\gamma$ and
Br$\delta$ are much weaker relative to H$_2$ lines. As the H$_2$ lines
are unlikely to change in time scales of a few years, this variability
in line ratios strongly suggests large fraction of the hydrogen
recombination line luminosity in near-IR is due to AGN illuminated gas
in NLR\@. The H$_2$ emission line fluxes are higher by $\sim$40 percent than
those reported by Vanzi et al. However, as discussed above, we were
forced to use the broad-band images of Vanzi et al.~for flux
calibration, and it is possible that this difference can at least
partly be attributed to changes in continuum brightness. This
interpretation is likely, as the $H$-band CO absorption lines are less
diluted by non-stellar continuum (Sect. \ref{sec_absorp}) than
detected by Vanzi et al.~(2002).

\begin{figure}
\includegraphics[width=8cm]{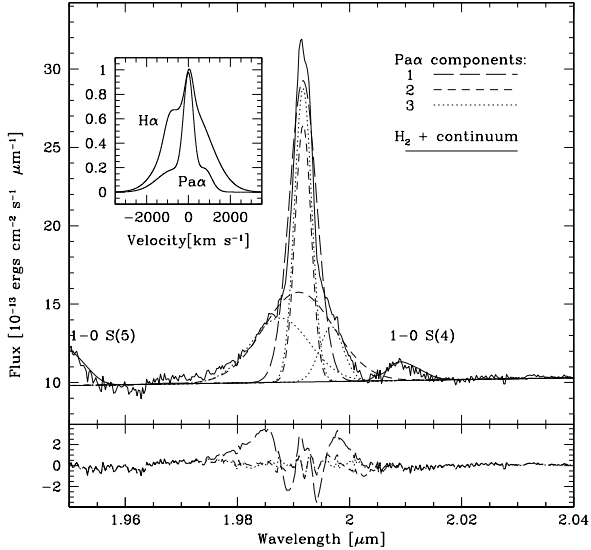}
\caption{The Pa$\alpha$ spectrum summed over the central 
0\farcs6$\times$0\farcs6. In the upper panel the fits with one 
(\textit{long-dashed}), two
(\textit{dashed}) and three Gaussians (\textit{dotted}) are indicated
as are the H$_2$ emission lines. In the lower panel
the residuals of the fits are shown.  The shape of the H$_2$ lines is
taken from the isolated 1--0 S(1) line. The inset shows the comparison
between the three component model of H$\alpha$ (Vanzi et al.~2002)
and Pa$\alpha$. \label{fig_pabroad}}
\end{figure}

A clear indication of an AGN in the Super-antennae is the presence of
coronal emission lines: {\sivi} (ionisation energy 167 eV) previously
detected by Vanzi et al.~(2002) and a faint {\alix} emission line (285
eV) detected at 12$\sigma$ level. While {\alix} has been detected only
in few AGN (Maiolino et al.~1999; Reunanen et al.~2003), the line is
one of the strongest NIR coronal lines when the ionising parameter
(the ratio of hydrogen ionising photon to total hydrogen densities) is
high enough. Detection of highly-ionised emission lines is not typical
in ULIRGs; Murphy et al.~(2001) found {\sivi} in only two galaxies out
of 33 ULIRGs in their survey. This simply reflects the fact that stars
are unable to photo-ionise the coronal emission lines, and the shocks
in starforming regions are not likely excitation mechanism
(e.g.~Marconi et al.~1994; Rodr\'igues-Ardila et al.~2002).

\begin{table}
\caption{Observed emission lines in the nucleus. Fluxes are given in
units of 10$^{-15}$ erg s$^{-1}$ cm$^{-2}$. {\iron} 1.64$\,\mu$m is
blended with Brackett 12-4 1.64117$\,\mu$m, {\iron} 1.68 $\mu$m with
Brackett 11-4 1.68112$\,\mu$m and Br$\delta$ with H$_2$ 2--1 S(5) 
1.94487$\,\mu$m.
}
\label{tab_nuc}
\begin{tabular}{@{}lccccc}
\hline
Line & $\lambda_0$ & $\lambda_{obs}$ & Flux & FWHM\\
& \AA & \AA &  &km s$^{-1}$\\
\hline
\iron                  & 15338.9 & 16314.3 $\pm$1.4 & 1.09  $\pm$ 0.06 & 1890 $\pm$ 80\\
\iron & 16439.9 & 17457.3 $\pm$0.4 & 5.10  $\pm$ 0.07 & 1110 $\pm$ 10\\
\iron & 16773.3 & 17826.4 $\pm$3.0 & 1.24  $\pm$ 0.10 & 1950 $\pm$ 130   \\
1--0 S(5)          & 18358.0 & 19472.2 $\pm$3.1  &  2.64 $\pm$ 0.4 & 1270 $\pm$ 160\\
Pa$\alpha$        & 18756.3 & 19909.8 $\pm$0.2  & 13.30 $\pm$ 0.04 &  515 $\pm$ 2\\
1--0 S(4)          & 18919.7 & 20090.8 $\pm$1.8  &  0.52 $\pm$ 0.05 &  830 $\pm$ 80\\
\nickel           & 19393.0 & 20586.9 $\pm$2.2  &  0.16 $\pm$ 0.04 &  550 $\pm$ 110\\
Br$\delta$
                  & 19451.0 & 20652.1 $\pm$1.1  &  0.37 $\pm$ 0.03 &  600 $\pm$ 40\\
1--0 S(3)          & 19575.6 & 20798.2 $\pm$0.3  &  3.34 $\pm$ 0.03 &  920 $\pm$ 10\\
\sivi             & 19634.1 & 20899.3 $\pm$0.9  &  0.72 $\pm$ 0.02 &  900 $\pm$ 30\\
1--0 S(2)          & 20337.6 & 21609.2 $\pm$0.3  &  1.20 $\pm$ 0.02 & 1140 $\pm$ 20\\
\alix             & 20449.9 & 21695.6 $\pm$1.3  &  0.12 $\pm$ 0.01 &  450 $\pm$ 50\\
\hei              & 20586.9 & 21863.7 $\pm$0.6  &  0.25 $\pm$ 0.01 &  550 $\pm$ 30\\
2--1 S(3)          & 20735.1 & 22020.4 $\pm$1.1  &  0.13 $\pm$ 0.02 &  500 $\pm$ 50\\
1--0 S(1)          & 21218.3 & 22536.5 $\pm$1.1  &  2.99 $\pm$ 0.02 &  953 $\pm$ 4\\
Br$\gamma$        & 21661.3 & 23002.3 $\pm$0.6  &  0.66 $\pm$ 0.02 &  540 $\pm$ 20\\
1--0 S(0)          & 22233.0 & 23607.7 $\pm$1.0  &  0.40 $\pm$ 0.03 &  580 $\pm$ 40\\
\hline
\end{tabular}
\end{table}

An intriguing similarity between our nuclear spectrum and that
presented in Vanzi et al.~(2002) is the presence of a faint emission
line at 2.058 $\mu$m, which they attributed to night sky OH emission
lines. However, the same feature is also present in our spectra which is
otherwise free of telluric night sky signatures.  The presence
of several, relatively broad emission lines (2--1 S(5) 1.94487$\,\mu$m,
Br$\delta$ 1.9451$\,\mu$m, 1--0 S(3) 1.95756$\,\mu$m, {\sivi} 1.96341$\,\mu$m) 
makes
the study of the region complicated. In order to reduce the number of free
parameters in the fit, we have assumed the line profiles of 1--0 S(3)
and Br$\delta$ are identical to those of the isolated 1--0 S(1) and 
Pa$\alpha$ lines,
respectively. Furthermore, we have used the 1--0 S(1) line profile for the 
2--1 S(5) line fitting, keeping the position of the line fixed relative to 
1--0 S(3) line.
The addition of the 2--1 S(5) line, which is strongly blended with Br$\delta$,
is necessary to make the residuals around Br$\delta$ smaller.  In
typical starforming galaxies with 1--0 S(1)/Br$\gamma$ $\sim$1
(e.g.~Goldader et al.~1997), the ratio 2--1 S(5) to Br$\delta$ is $\sim$0.06
assuming thermal excitation with $T_{ex} = 2000$~K\@. However, due to the
rather extreme 1--0 S(1)/Br$\gamma$ ratio of $\sim$5 observed in the
Super-antennae, the contribution of 2--1 S(5) is $\sim$30 percent of
Br$\delta$ flux. The inclusion of 2--1 S(5) does not affect the
position of the unidentified line. The fitting gives the wavelength
for the unidentified line 2.0584$\pm$0.0001 $\mu$m ($\lambda_0 \simeq
1.937$ $\mu$m).  The best candidate in the NIST Atomic Spectra Database is
{\nickel} 19393{\AA} corresponding to the 
3d$^8$~4s~a$^4$F--3d$^8$~4s~a$^2$F transition. Optical emission lines 
($\lambda\lambda$ 7380 and
7414) originating from the same upper energy level have been detected
in objects of various types, including nebulae (e.g.~Osterbrock, Tran
\& Veilleux 1992), supernova remnants (e.g.~Dennefeld 1986) and AGN
(e.g.~Halpern \& Oke 1986). As the 1.93 $\mu$m line should also be
relatively strong (based on the relative $A$-values), the
identification of this line as {\nickel} seems secure. The optical
lines have, however, not been detected in the Super-antennae (Berta et
al.~2003).

\subsection{Emission line profiles}

The most notable feature of the nuclear spectrum is of course
Pa$\alpha$. However, Pa$\alpha$ is badly fitted with a single Gaussian
(Fig. \ref{fig_pabroad}). Similarly, multiple components (three or
more; Vanzi et al.~2002; Colina, Lipari \& Macchetto 1991) have been
used to approximate the optical emission lines. The wavelength region
($\sim$1.99 $\mu$m) around Pa$\alpha$ appears to be
completely free of atmospheric (telluric) residuals - redward of
Pa$\alpha$ the strength, shape and position of 1--0 S(4) at 1.89
$\mu$m is in excellent agreement with the measurements of the
stronger and isolated 1--0 S(1) line. Therefore, the continuum level
around P$\alpha$ line is very well determined. Pa$\alpha$ is
asymmetric, with the blue wing being stronger, and is similar to the
H$\alpha$ profile observed by Vanzi et al.  While a two-component fit
produces acceptable results in the sense that the residuals (red
asymmetry) can be readily explained by obscured broad line region, the
similarity with the H$\alpha$ suggests this is not the case. However,
we note that the FWHM of the broader component in the two-component
fit is similar to possible detection of broad H$\alpha$ in polarised
emission by Pernechele et al.~(2003).

When compared to H$\alpha$, the center-most component in Pa$\alpha$ is
much stronger and represents $\sim$50 percent of the total Pa$\alpha$
emission, while the narrow ($\sigma=185$ km s$^{-1}$) component of
H$\alpha$ provides only 6.5 percent of the total line emission (Vanzi et al
2002). Qualitatively the weakness of coronal emission lines relative
to the hydrogen recombination lines when compared with other Sy2
galaxies suggest the central component arises from a starburst and not
from the NLR.

All the three main emission lines (Pa$\alpha$, H$_2$ 1--0 S(1) and
\iron) have different line profiles. H$_2$ has a strong red component,
while {\iron} has a weak asymmetry to the red and is broad. Since
{\iron} has critical density of the order of $N_e\sim10^4 - 10^5$
cm$^{-3}$ (Blietz et al. 1994), the BLR cannot contribute
significantly to the flux. Thus the shape of {\iron} also supports
multiple velocity components instead of a genuine broad component.
Careful deblending also suggests that {\sivi} may be double-peaked,
with a separation of $\sim$750 km s$^{-1}$ ($z=0.06155, 0.06419$)
between the two peaks, indicating it may originate from an unresolved
disk. {\alix} is too weak to determine whether the profile is similar
to {\sivi}.

\subsection{Nuclear absorption lines\label{sec_absorp}}

In addition to many prominent emission lines, the nuclear spectrum
also contains several absorption lines, allowing the application of
the penalised pixel fitting method (Cappellari \& Emsellem 2004) to 
determine the line of sight velocity distribution.
In this process, the line of sight velocity
distribution is described by a Gauss-Hermite series (van der Marel \&
Franx 1993) and the solution is based on the maximum penalised
likelihood formalism (see Cappellari \& Emsellem 2004 and references
therein for details).  The engineering grade detector used in SINFONI
during the time of the observations had a large dead region at the red
end of the spectrum, which compromised the data quality in the
$K$-band beyond $>$2.4 $\mu$m. Therefore, only the $H$-band was
used. The fitting was done using only the low-order moments ($v$,
$\sigma$), as including the higher orders $h_3$ and $h_4$ did not
provide a significant improvement in the quality of the fit, most
likely due to relatively low S/N and beam-smearing. Late type stars
(K3V, M0V, M5III)
obtained with SINFONI for another observing program were used as
templates.

We derive $\sigma$ = 210 $\pm$ 20 km s$^{-1}$, in agreement 
with $\sigma$ = 175 $\pm$
24 km $^{-1}$ of Dasyra et al.~(2006) and 188 $\pm$ 10 km $^{-1}$ of
Tacconi et al.~(2002). The H-band continuum is consistent with late
type stellar template, with no need to introduce a non-stellar
component to the fit. This is in contrast to the findings of Vanzi
et al.~(2002),
according to whom the non-stellar component contributed 30 percent of the
$H$-band emission based on a similar fitting on the $H$-band and an even
higher fraction based on broad-band colours.

The disappearance of the non-stellar component is not surprising as our
observations and those of Vanzi et al.~were separated by 5 years. The
most likely origin of the diluting continuum is hot dust heated close
to the sublimation temperature, and therefore follows closely changes
in the heating flux. Alternatively, the diluting continuum may be due
to synchrotron, as has been suggested for some of the nearby AGN in
near-IR (Prieto et al., in preparation). In both cases the
implication is that, at least in near-IR, the non-stellar continuum is
solely due to the AGN\@. As discussed above, the Super-antennae may be AGN
dominated even in mid-IR (Charmandaris et al.~2002; Risaliti et
al.~2003; 2006).

\subsection{Emission line morphologies}

\begin{figure*}
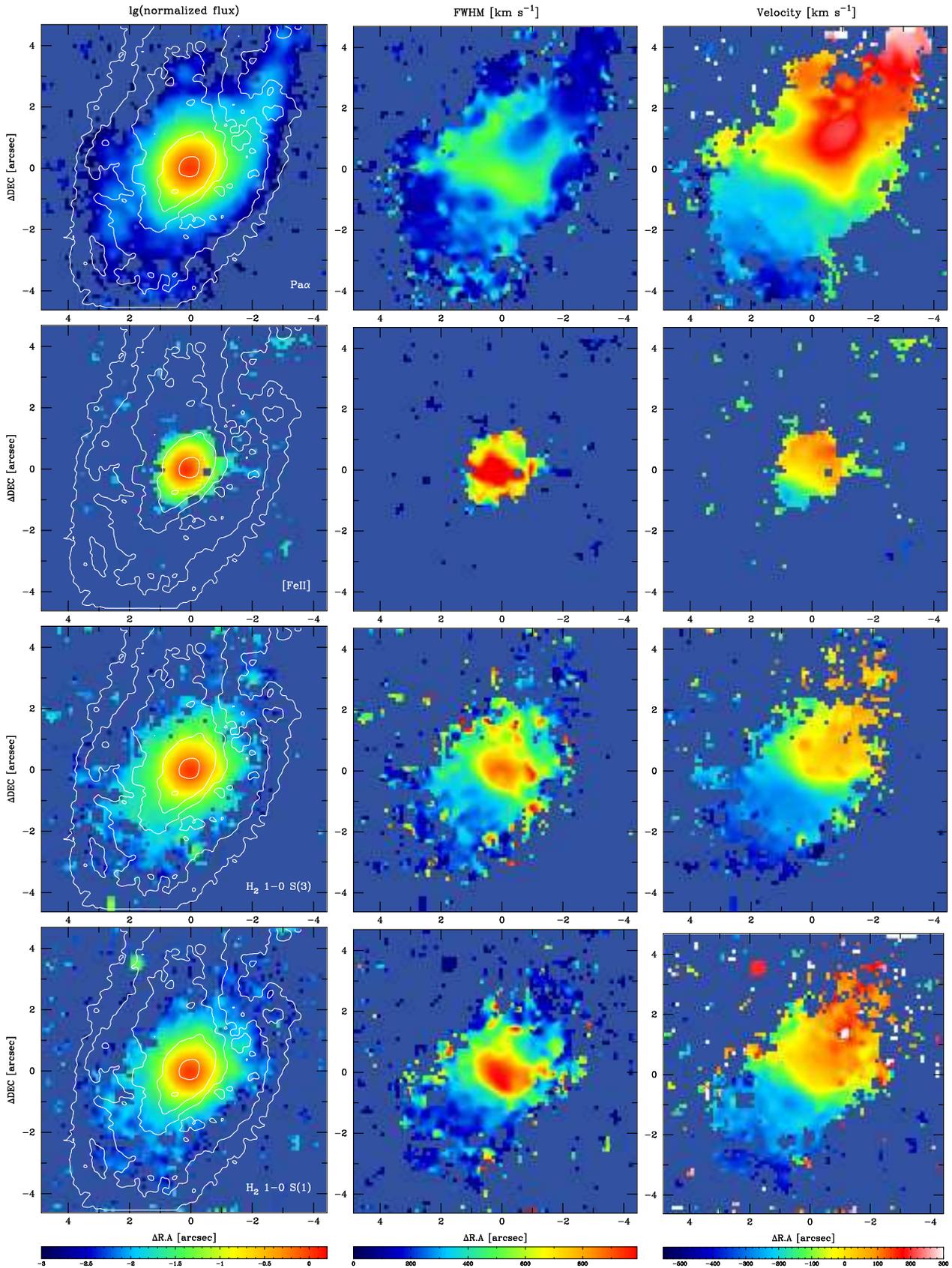

\includegraphics[width=5.72cm]{pa_flux_color_low.ps}
\includegraphics[width=5.4cm]{pa_fwhm_color_low.ps}
\includegraphics[width=5.4cm]{pa_velocity_color_low.ps}

\includegraphics[width=5.72cm]{feii_flux_color_low.ps}
\includegraphics[width=5.4cm]{feii_fwhm_color_low.ps}
\includegraphics[width=5.4cm]{feii_velocity_color_low.ps}

\includegraphics[width=5.72cm]{h2s3_flux_color_low.ps}
\includegraphics[width=5.4cm]{h2s3_fwhm_color_low.ps}
\includegraphics[width=5.4cm]{h2s3_velocity_color_low.ps}

\includegraphics[width=5.72cm]{h2s1_flux_color_low.ps}
\includegraphics[width=5.4cm]{h2s1_fwhm_color_low.ps}
\includegraphics[width=5.4cm]{h2s1_velocity_color_low.ps}
\caption{From left to right, the flux normalised to the peak nuclear
flux, Gaussian FWHM and the velocity field for Pa$\alpha$ ({\em top
row}; excluding the ``broad'' component), {\iron} ({\em middle row})
and H$_2$ 1--0 S(1) ({\em bottom row}). The optical morphology from
HST 0.8 $\mu$m image is shown in contours. \label{fig_many}}
\end{figure*}

The emission line fluxes relative to the nuclear peak flux, the full
width at half-maximum (FWHM) of the Gaussian fitted to the line
profile and the velocity field for Pa$\alpha$ 1.87 $\mu$m, {\iron}
1.64 $\mu$m and the H$_2$ line 1--0 S(1) 2.12 $\mu$m are displayed in
Fig. \ref{fig_many}. For comparison, the HST 0.8 $\mu$m image mildly
smoothed to 0\farcs3 resolution is shown in contours. The alignment
between the HST image and SINFONI data was based on the position of
the nuclear continuum source. While the southern nucleus is red, and
the optical nucleus may therefore not coincide with the near-IR
nucleus, no evidence for this is found based on comparison between
large-field near-IR and HST images, at least within the
seeing-limited resolution of data discussed in this paper. In case of
Pa$\alpha$, the continuum wavelength range in automatic line fitting
has been selected to exclude contribution from ``broad'' component;
thus the fitting procedure (and line ratios) is comparable with the other
lines where the Pa$\alpha$ components cannot be detected.

As the SINFONI field of view does not contain any stars, a direct
estimate for the achieved spatial resolution cannot be derived based
on the data itself. Furthermore, the DIMM sensor measures optical
seeing close to zenith while the Super-antennae was observed at an airmass
of $\sim$1.5. The nucleus is unresolved in the $K$-band image (Vanzi
et al.), where the spatial resolution $\sim$1\farcs0 (1.3 kpc) can be
derived directly from the numerous stars present in the large
f.o.v.  After collapsing the spectral cubes in emission line-free
wavelengths, we derive the FWHM size for the nuclear continuum source
0\farcs9 both in $H$ and $K$, similar to that measured from pure emission
line maps for the main emission lines (Pa$\alpha$, {\iron}, H$_2$
lines). While these values are larger than reported by the DIMM in the
optical, especially during $H$-band observations, the derived nuclear
sizes are inconsistent with the unresolved nucleus in broadband
images. More likely, the DIMM sensor has underestimated the optical
seeing and the nucleus is unresolved in both the collapsed spectral
continuum images and emission lines.

While the nuclear source itself is unresolved, a fainter extended
component can be detected in most emission lines. The most extended
emission line is Pa$\alpha$ 1.8756 $\mu$m, which is dominated by the
nucleus with spiral structures extending to south-east and
north-west. These Pa$\alpha$ features coincide with the dusty spiral
present in HST archive images (Fig. \ref{fig_f804w}).  Similar to
H$\alpha$ emission (Mihos \& Bothun 1998), both Pa$\alpha$ and
molecular H$_2$ lines show enhanced emission northwest of the nucleus.
The southern spiral consisting of several clumps (starforming regions)
is also faintly traced in H$_2$. Furthermore, a separate starforming
clump is seen only in Pa$\alpha$ $\sim$3{\arcsec} north of the nucleus
coinciding with the bridge detected in continuum
(Fig. \ref{fig_f804w}) connecting the southern galaxy to the northern.

\subsection{Molecular gas}

Davies
et al.~(2003) studied the molecular gas in 7 ULIRGS and found that
while the 1--0 transitions appear to be thermalised, the same is not
true for 2--1 and 3--2 transitions. They concluded the H$_2$ excitation
is predominantly UV fluorescence, but the lower levels are thermalised
by inter-molecule collisions in dense clouds. This produces different
excitation temperatures, $\sim$1300 K based on $v = 1$ transitions and
$\sim$5000-6000 K based on $v = 2$ and $v = 3$ transitions.  In order
to probe the fainter transitions, we have corrected the spectra for
rotation with velocities derived from the 1--0 S(1) line. This ``derotation''
increases S/N of the spectra by decreasing the width of the spectra
features and decorrelates possible telluric residuals; this procedure
is identical to what was used in Davies et al.~(2005).  However, due to a 
lack of
CO-bandheads at $K$-band, the template fitting is so badly constrained,
that we are unable to remove the stellar contribution.

\begin{table}
\caption{Comparison between the H$_2$ emission line ratios in ULIRGs,
AGN and the Super-antennae; the Br$\gamma$ for the extended case is estimated
from Pa$\alpha$\label{table_comp}}
\begin{tabular}{lcccc}
\hline
Line     & ULIRG           & AGN             & \multicolumn{2}{c}{Super-antennae} \\
         &                 &                 & Nucleus & Extended \\
\hline
1--0 S(3) & ...             & ....            & 0.595  & 0.739 \\
1--0 S(2) & 0.337$\pm$0.018 & 0.379$\pm$0.068 & 0.496  & 0.327\\
2--1 S(3) & 0.146$\pm$0.022 & 0.217$\pm$0.132 & 0.045  & 0.045\\
1--0 S(1) & 1.000           & 1.000           & 1.000  & 1.000\\
2--1 S(2) & 0.074$\pm$0.007 & 0.118$\pm$0.012 & 0.039  & $<0.06$\\
1--0 S(0) & 0.271$\pm$0.035 & 0.327$\pm$0.047 & 0.308& 0.350\\
2--1 S(1) & 0.142$\pm$0.031 & 0.125$\pm$0.038 & ...    & ...\\
Br$\gamma$ & 0.6--1.33   &    & 0.256   & 0.164 \\
\hline
\end{tabular}
\end{table}

\begin{figure}
\includegraphics[width=8cm]{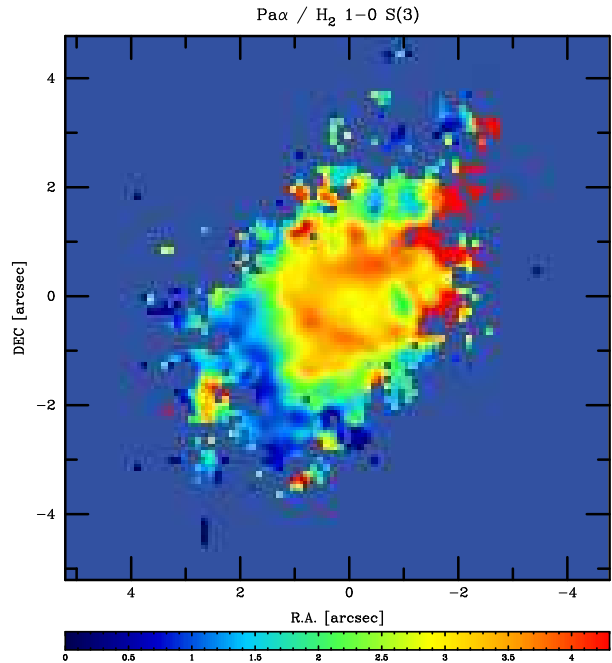}
\caption{The Pa$\alpha$/H$_2$ 1--0 S(3) ratio
\label{fig_pa_per_10s3}}
\end{figure}

The comparison between the average line ratios presented by Davies et
al.~(2003) for ULIRGS, Davies et al.~(2006) for AGN and the Super-antennae
is presented in Table \ref{table_comp}. For the Super-antennae, the line
ratios both for the nuclear region ($r<$0\farcs6) and extended
emission ($1\farcs25<r<2\farcs5$) are given. When compared with the ULIRG
sample, the 2--1 transitions appear weaker both at the nucleus and in
the extended annulus, and are not overpopulated relative to the 1-0
transitions.
Br$\gamma$/1--0S (1) is much lower at the nucleus, and in the extended
annulus Br$\gamma$ cannot be even detected.  Assuming the Case B
Pa$\alpha$/Br$\gamma$ ratio of 12.1, we infer a  Br$\gamma$/1--0 S(1)
ratio of $\sim$0.16 there. The decrease in Pa$\alpha$/1--0 S(3) ratio
is also clearly visible in Fig. \ref{fig_pa_per_10s3}, which
interestingly also reveals a small ring-like structure with diameter
$\sim$1\farcs5.

\begin{figure}
\includegraphics[width=8cm]{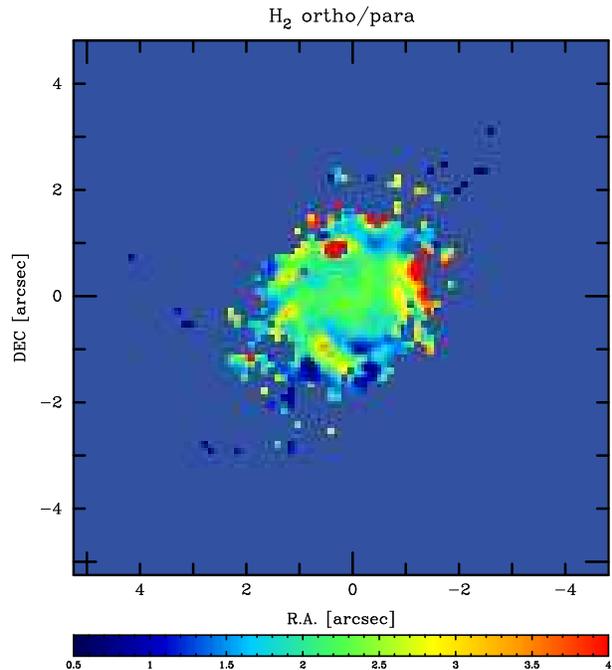}
\caption{Ortho-Para ratio of H$_2$ determined from 1--0 lines S(1), S(2), S(3)
\label{fig_ortho}}
\end{figure}

We have estimated the excitation temperature from 1--0 S(3)/1--0
S(1), and ortho/para ratio from 1--0 S(2)/1--0 S(1) since these are
the strongest H$_2$ emission lines in the spectra. The ortho/para
ratio can easily be shown to be practically independent of
extinction:
\begin{equation}
f_{o/p} = \frac{1.1102 \left[S(3)/S(1)\right]^{0.449}}{S
(2) / S(1)} 10^{-0.004\,A_K}, 
\end{equation}
where $S(x)$ denotes the observed intensities for 1--0 S(x) emission
lines and $A_K$ is the $K$-band extinction assuming $A_\lambda \propto
\lambda^{-1.85}$ and the $A$-coefficients from Wolniewicz, Simbotin \&
Dalgarno (1998).
The ortho/para map is shown in Fig. \ref{fig_ortho}, showing a
relatively constant ratio. By summing the ``derotated'' spectra in
annuli, we can follow the ortho/para ratio changes further out than
calculating it in individual spectra. At the nucleus ($r<0\farcs6$)
the ratio is 2.4$\pm$0.1, and remains similar in the annulus
$0.6<r<1.25\arcsec$. Further out, the ratios are consistent with
thermal excitation of H$_2$ (ortho/para $\sim$3). As the minimum
values detected in ortho/para ratio coincide with the ring-like
structure visible in Pa$\alpha$/1--0 S(3) ratio, it is likely that in
the nuclear region H$_2$ is excited by UV fluorescence in dense
clouds. A contribution from X-ray excitation may be needed to
depopulate the 2--1 states (e.g.~Gredel \& Dalgarno 1995), as Davies et
al.~(2003) observed larger ratios in their ULIRG sample. Similar
scenario has been suggested for NGC\,6240 (Draine \& Woods 1990),
though higher-resolution data is clearly needed to properly resolve
the possible starbursting ring and separate the mechanisms.

For the extended H$_2$ gas the situation is more clear, as both the
ortho-para ratio and low Pa$\alpha$/H$_2$ line ratio suggest thermal
excitation through shocks. While in principle X-rays can also produce
low Pa$\alpha$/H$_2$ ratios (Gredel \& Dalgarno 1995), such a scenario
seems unlikely, as the likely origin for the extended X-ray emission
is supernovae; however, {\iron} is strongly concentrated at the
nuclear region.
 
The average column density of molecular hydrogen within the 5$\times5$
pixel region is $1.8\times 10^{18}$\,cm$^{-2}$, corresponding to the total
mass of hot excited molecular hydrogen of $1.8\times 10^4\,$M$_\odot$. The peak
(seeing dependent) column density is a factor of 1.3 bigger than the
average value. The total mass of the available molecular material as
determined from CO radio emission is several orders of magnitudes
larger, 3$\times$10$^{10}$ M$_\odot$ (Mirabel et al.~1990); however,
the radio beamsize (44\arcsec) includes both the nuclei in the system.

\subsection{Kinematics}

Low resolution H$\alpha$ velocity maps for the Super-antennae were
presented by Mihos \& Bothun (1998), whose Fabry-Perot images also
covered the northern galaxy. The most notable difference between our
maps and theirs is the plume detected in H$\alpha$ in East of the
nucleus, which is not very extended in Pa$\alpha$. This may be simply
due to lower signal to noise ratio at the outer edges of the mosaiced
spectra. The faint streaming motion with $\Delta v \simeq -130$
km$^{-1}$ appears to be present in Pa$\alpha$ in South of the nucleus
extending up to 4{\arcsec} (Fig. \ref{fig_many}).

\begin{figure}
\includegraphics[width=8.3cm]{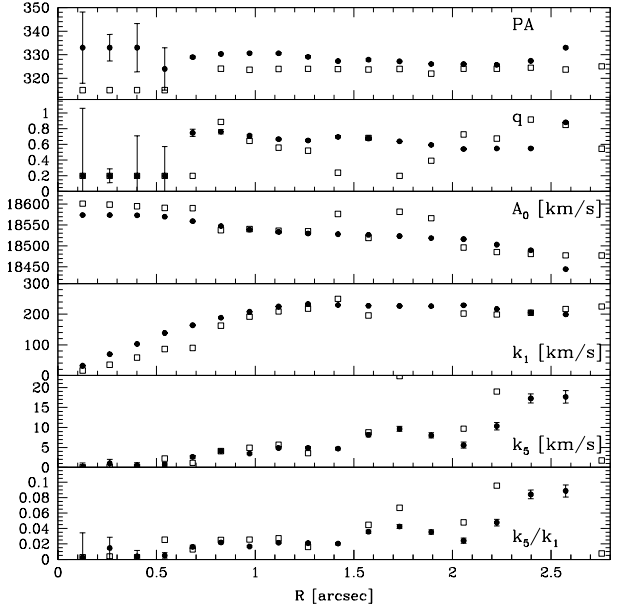}
\caption{Kinematic model parameters for the Pa$\alpha$ emission
(\textit{circles with error bars}) and H$_2$ 1--0 S(1) (\textit{open
squares}). From top to bottom: the position angle of the ellipses, the
flattening coefficient $q$ (related to $\epsilon$ by $q=1-\epsilon$),
the systemic velocity $A_0$, rotational velocity $k_1$ and
non-rotational velocity component $k_5$. \label{fig_model_paras}}
\end{figure}

We have applied the Kinemetry IDL routine described by Krajnovi\'c et
al.~(2006) to the data. This method is a generalisation of surface
photometry performed in isophotal annuli of galaxies applied to
higher order moments ($v$, $\sigma$, $h_3$, $h_4$...). The routine can
either use a constant systemic velocity or allow it to change as a
function of radius; for isolated, non-interacting galaxies the
systemic velocity ($A_0$ following the terminology in Krajnovi\'c et
al.~2006) is expected to be nearly constant. We have tried both
methods with Pa$\alpha$ in the Super-antennae, but constant $A_0$ produces a
poorer match with the observed velocity field. The parameters of the
warped disk model with $A_0$ kept as a free parameter are shown in
Fig. \ref{fig_model_paras}. For the four innermost radii, the fitting
is done inside the PSF and therefore the position angle and
ellipticity have very large errors. This simply represents the fact
that while the fitting process is numerically stable even inside the
HWHM radius of the PSF, the shape parameters derived there have no
real physical meaning.

The fitting has revealed the presence of at least two separate
kinematic components within the central $r=4$\arcsec in the Super-antennae,
possibly three. In the inner parts of the galaxy the velocity field is
well modelled by a warped disk model, and the residuals between the
observed velocities and model are generally within $\sim$25 km
s$^{-1}$ . Northeast and west of the nucleus, the residuals are much larger, up to
150 km s$^{-1}$. While the shape parameters ($q$ and position angle)
of the \textit{outermost} ellipses does depend on how large the area
the fitting is carried out over or whether the beginning of the spirals
arms is included or not, the closeness of the edge of usable data does
not affect these residuals significantly. Especially northeast of the
nucleus, the area of large residuals is also clearly visible in the
observed velocity field (Fig. \ref{fig_many}).

\begin{figure*}
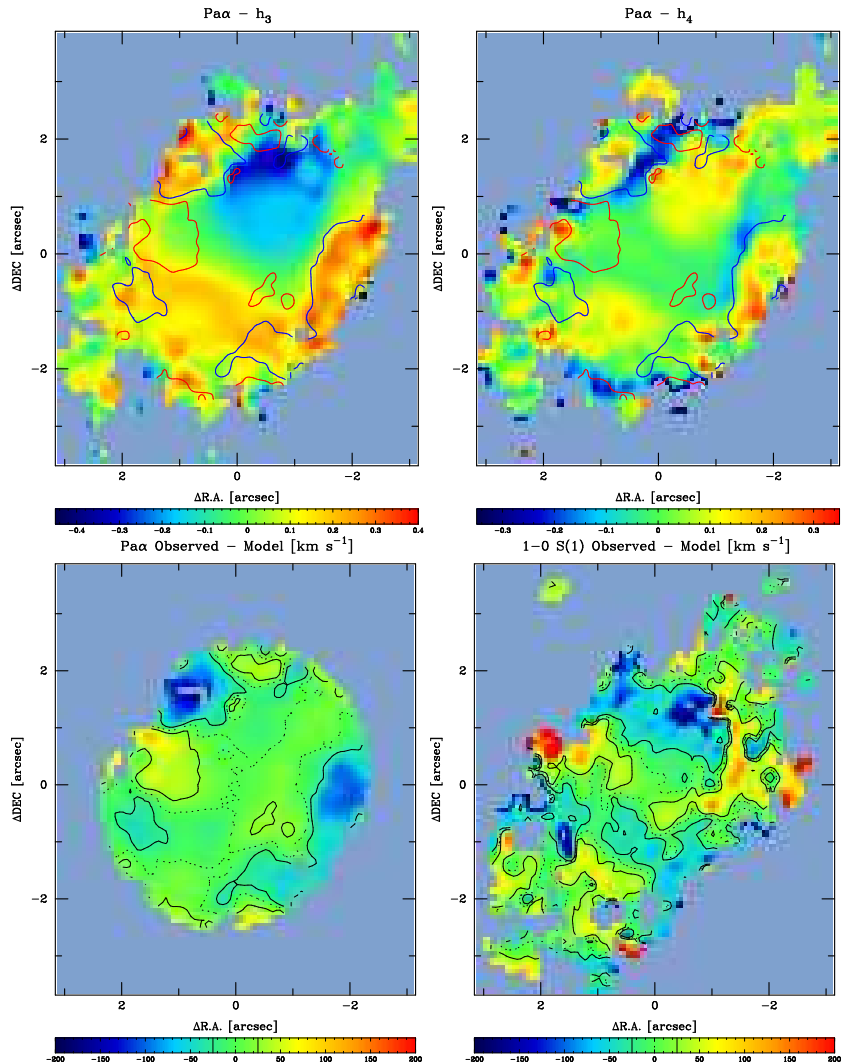

\includegraphics[height=7cm]{h3_low.ps}
\includegraphics[height=7cm]{h4_low.ps}

\includegraphics[height=7cm]{ero_low.ps}
\includegraphics[height=7cm]{ero_h2_low.ps}
\caption{Kinematic model of Pa$\alpha$ and 1--0 S(1) emission in the
Super-antennae. The $h_3$ and $h_4$ components from Gauss-Hermite
expansion of the Pa$\alpha$ profile are shown \textit{upper left} and
\textit{upper right}, respectively.  The difference between $v_{obs}$
and $v_{model}$ constructed without including higher order harmonics
(rotation only) is shown for Pa$\alpha$ (\textit{lower left}) and 1--0
S(1) (\textit{lower right}). The contours in $h_3$ and $h_4$ figures show the
Pa$\alpha$ regions with $v_{obs}$ - $v_{model}$ = $\pm$25 km
s$^{-1}$. \label{fig_model} }
\end{figure*}

The ellipticity of the fitted rings, related to the flattening
parameter of Krajnovi\'c et al.~(2006) by $q=1-\epsilon$, gradually
increases from 0.24 at $r=0\farcs68$ to 0.45 at $r>2$\arcsec, while
the position angles of the ellipses are almost constant, 
327--331\degr. $A_0$ represents the systemic velocity of the fitted
ellipses and is relatively stable between 0\farcs8 -- 2\arcsec,
$\sim$18530 km s$^{-1}$. Further out, $A_0$ decreases and the ratio
$k_5$ increases. As the presence of multiple components is detected by
the rise of the $k_5$ coefficient in parametrisation of Krajnovi\'c et
al.~(2006), this indicates that the outer parts of the galaxy do not
follow the same disk-like kinematics as the inner parts. This outer
component is already clearly visible in the observed velocity field
(Fig. \ref{fig_many}). Further support for the separate outer
component comes from the Gauss-Hermite expansion of the Pa$\alpha$
profile (Fig. \ref{fig_model}), where the higher order moments $h_3$
and $h_4$ are directly observable quantities without any modelling
assumptions. $h_3$ and $h_4$ in the inner regions closely follow the
expectations for a warped disk located at position angle of
$\sim$330\degr. Northeast and west of the nucleus, both $h_3$ and $h_4$ show
significant deviations from disk-like rotation; these deviations
coincide with the the regions with $v_{obs} - v_{model} < -25$ km
s$^{-1}$ of the model fitting.

Closer still to the nucleus, with $r<0\farcs7$, the fitted $A_0$
values are higher than in the surrounding region. The nature of this
jump is not clear based on the current, seeing limited data. The
redshifted nuclear velocities appear to be connected with the positive
residuals present in $v_{obs} - v_{model}$ map at position angle of
$\sim$70\degr (Fig. \ref{fig_model}).

We have repeated an identical fitting procedure with 1--0 S(1) as we
did with Pa$\alpha$. Overall, the quality of the kinemetric fitting on
1--0 S(1) is limited both by lower S/N and smaller spatial extent of
the data.  The most notable differences between the velocity field of
molecular gas and ionized medium is that the 1--0 S(1) is redshifted
by 80 km s$^{-1}$ relative to Pa$\alpha$ at the nucleus and
blueshifted by $\sim$90 km s$^{-1}$ in the Northwest. Despite these
differences, the H$_2$ velocity field is consistent with rotating disk
with similar kinemetric parameters than derived for Pa$\alpha$
(Fig. \ref{fig_model_paras}). Relatively large $v_{obs} - v_{model}$
residuals upto $\sim$-100 km s$^{-1}$ are detected Northwest of the
nucleus coinciding with the region where 1--0 S(1) is blueshifted
relative to Pa$\alpha$ (Fig. \ref{fig_model}). Interestingly, 1--0
S(3) velocity field is significantly different from 1--0 S(1)
northwest of the nucleus. The blue component of S(3) emission line is
stronger than in S(1); thus the excitation temperature of the blue
component is higher. This difference is unlikely to be due to
extinction, as the wavelength difference between the lines is
relatively small and therefore they both suffer from similar amounts
of reddening.

\section{Conclusions}

We have presented the results of near-IR integral field
spectroscopy of the southern galaxy of the Super-antennae. The main
results obtained from this project are:
\begin{itemize}
\item{Pa$\alpha$ arises in a warped disc with position angle of
$\sim$330\degr and inclination $i$=40--55\degr. The kinemetric
parameters derived for H$_2$ are similar.}
\item{Unlike in previous studies, we found the $H$-band continuum
to be consistent with pure stellar emission, with no evidence for
diluting non-stellar continuum at present.}
\item{The Pa$\alpha$/H$_2$ line ratio maps reveal the presence of the
ring-like structure with diameter $\sim$1.5--2.0\arcsec, coinciding
the the region with lowest H$_2$ ortho/para ratios. This morphological
argument together with non-thermal ortho-para ratios suggest UV
fluorescence in dense clouds is an important H$_2$ excitation mechanism
in the nuclear region.}
\item{Extended H$_2$ emission with $r>1\farcs3$ has thermal ortho/para
ratio and very low Pa$\alpha$/H$_2$ line ratios, indicating that thermal
excitation in shocks is important at larger radii.}
\end{itemize}

\section{Acknowledgements}

We thank Leonardo Vanzi for kindly providing us the broadband images
and Richard McDermid for fruitful discussion during the work. We are
grateful to Davor Krajnovi\'c and Michele Cappellari for making their
software publicly available. This work is based on observations
collected at the European Southern Observatory, Paranal, Chile,
(60.A-9041(A)). This research has made use of the NASA/IPAC
Extragalactic Database (NED) which is operated by the Jet Propulsion
Laboratory, California Institute of Technology, under contract with
the National Aeronautics and Space Administration.

\label{lastpage}

\end{document}